\documentclass{nature}

\usepackage{amsmath}
\usepackage{graphicx}
\usepackage{xfrac}
\usepackage{multirow}
\usepackage{subfigure}
\usepackage{mathtools}
\usepackage[colorlinks=true,linkcolor=blue,citecolor=black,urlcolor=blue]{hyperref}
\usepackage{soul}
\usepackage[usenames,dvipsnames]{xcolor}
\usepackage{authblk}
\usepackage[labelfont=bf]{caption}
\usepackage[figurename=Figure ]{caption}
\usepackage{siunitx}
\usepackage{scalefnt}

\makeatletter

\let\saved@includegraphics\includegraphics
\AtBeginDocument{\let\includegraphics\saved@includegraphics}

\makeatother

\usepackage{color}
\usepackage{dsfont}
\usepackage[normalem]{ulem}
\usepackage{dcolumn}
\usepackage{blindtext}
\usepackage{outlines}
\usepackage{enumitem}
\usepackage{soul}
\setlist[enumerate,2]{label=\roman*)}
\setlist[enumerate,3]{label=\alph*)}
\usepackage{multirow}

\newcommand{\onlinecite}[1]{\hspace{-1 ex} \nocite{#1}\citenum{#1}} 
\newcommand{\VEC}[1]{\mathbf{#1}}

\begin{document}

\title{Long range and highly tunable interaction between local spins coupled to a superconducting condensate}

\author{Felix K{\"u}ster$^{1}$, Sascha Brinker$^{2}$, Samir Lounis$^{2,3\ast}$, Stuart S. P. Parkin$^{1\ast}$, Paolo Sessi$^{1\ast}$}
\affil{$^1$Max Planck Institute of Microstructure Physics, Halle 06120, Germany\\
$^2$Peter Gr{\"u}nberg Institut and Institute for Advanced Simulation, Forschungszentrum J{\"u}lich \& JARA, J{\"u}lich D-52425, Germany\\
$^3$Faculty of Physics, University of Duisburg-Essen and CENIDE, 47053 Duisburg, Germany}
\affil{$^\ast$ Emails: s.lounis@fz-juelich.de, stuart.parkin@mpi-halle.mpg.de, paolo.sessi@mpi-halle.mpg.de}

\date{}

\maketitle

\begin{abstract}

Interfacing magnetism with superconducting condensates is rapidly emerging as a viable route for the development of innovative quantum technologies. In this context, the development of rational design strategies to controllably tune the interaction between magnetic moments is crucial. In the metallic regime, the indirect interaction mediated by conduction electrons, the so-called RKKY coupling, has been proven to be remarkably fertile in creating and controlling magnetic phenomena. However, despite its potential, the possibility of using superconductivity to control the sign and the strength of indirect interactions between magnet moments remains largely unexplored. 
Here we address this problem at its ultimate limit, demonstrating the possibility of maximally tuning the  interaction between local spins coupled through a superconducting condensate with atomic scale precision. By using Cr atoms coupled to superconducting Nb as a prototypical system, we use atomic manipulation techniques to precisely control the relative distance between local spins along different crystallographic directions while simultaneously sensing their coupling by scanning tunneling spectroscopy. Our results reveal the existence of highly anisotropic superconductor-mediated indirect couplings between the local spins, lasting up to very long distances, up to 12 times the lattice constant of Nb. Moreover, we demonstrate the possibility of controllably crossing a quantum phase transition by acting on the direction and interatomic distance between spins. The extremely high tunability provides novel opportunities for the realization of exotic phenomena such as topological superconductivity and the rational design of magneto-superconducting interfaces.

\end{abstract}


\section*{Introduction}
Interfacing materials with radically different properties is a viable route to discover new physical phenomena and create new functionalities going beyond those hosted by each one of the building blocks \cite{OH2004,Wang_2012,Zhonge1603113}.  
In this context, the interface between a superconductor and a magnet is one of the most remarkable examples \cite{LR2015}. 
These ordered phases have been for a very long time thought to be mutually exclusive \cite{PhysRev.114.977,PhysRevLett.1.92,PhysRevLett.9.315}. 
However, over the last decade, it has been experimentally demonstrated that under the right conditions, at a superconductor–magnetic interface both superconductivity and spin polarization can coexist creating new states such as Majorana modes \cite{Nadj-Perge602,PhysRevLett.115.197204} and spin-triplet Cooper pairs \cite{Robinson59,DDG2015}, both offering tantalizing opportunities for radically new technologies .

Accelerating the progress in the field crucially relies on identifying, understanding, and controlling the interaction between superconducting condensates and magnetic order. 
In this context, indirect magnetic coupling, i.e. the coupling of magnetic moments mediated via the superconducting condensate, occupies a central role. 
In conventional metals, the polarized spins of conduction electrons can create such links called Ruderman-Kittel-Kasuya-Yosida (RKKY) interactions\cite{Ruderman1954,Kasuya1956,Yosida1957}. 
Their study proved to be essential to the development of spin-valve and tunneling magnetoresistance devices, through spin-engineered and especially synthetic antiferromagnet hetero-structures that have had a tremendous impact on the exponential increase in magnetic storage data capacity \cite{Parkin.1990,PhysRevLett.67.3598,PhysRevB.44.7131}.
With the advancements in scanning tunneling microscopy (STM) and spectroscopy (STS) techniques it became possible to scrutinize RKKY interactions on the nanometer scale by manipulation of magnetic adatoms on a metal substrate\cite{Eigler}, thereby creating atomically precise spatial maps of the coupling strength up to \SI{1.5}{\nano\meter} between the atoms\cite{Zhou.2010,Khajetoorians.2012,Khajetoorians2016}.

In superconducting condensates, it was theoretically predicted that besides the RKKY interactions, additional longer range spin coupling can be enabled by Cooper pairs with an exponential decay over the superconducting coherence length and which can dominate over conventional RKKY even at distances significantly smaller than the coherence length~\cite{Kochelaev1979,PhysRevLett.113.087202}. Despite their relevance for both fundamental aspects as well for their far reaching implications for the design and creation of non-trivial magnetosupeconducting states of matter, the indirect coupling between local magnetic moments mediated by superconducting condensates remain largely unexplored.

Here, we combine atomic manipulation techniques with high resolution spectroscopy to investigate the emergence of superconducting condensate-mediated coupling between magnetic moments at the most fundamental level, i.e. the atomic scale. 
By systematically sensing the interaction between localized spins placed at different distances and along different crystallographic directions, we demonstrate that: (i) local spins can talk at large distances, up to several lattice constants, (ii) the use of a superconductor hosting a highly anisotropic Fermi surface allows to precisely tune the interaction depending on the crystallographic direction, and (iii) the possibility to tailor not only the strength, but also the sign of the energy shift induced by the interaction, and to use it to controllably cross the quantum phase transition (QPT). The latter is obtained in two distinct ways: by changing the crystallographic direction and/or the distance between magnetic impurities. A direct comparison with interactions mediated by conventional conduction electrons confirms the high anisotropy of the coupling and the existence of longer range effects in the superconducting regime. Our findings project a plethora of opportunities for highly controllable magneto-superconducting interactions that can be extended to more complex systems.

\section*{Results}
\subsection{Experimental lineup.}

\begin{figure*}
        \renewcommand{\figurename}{Figure}
	\centering
	\includegraphics[width=.95\textwidth]{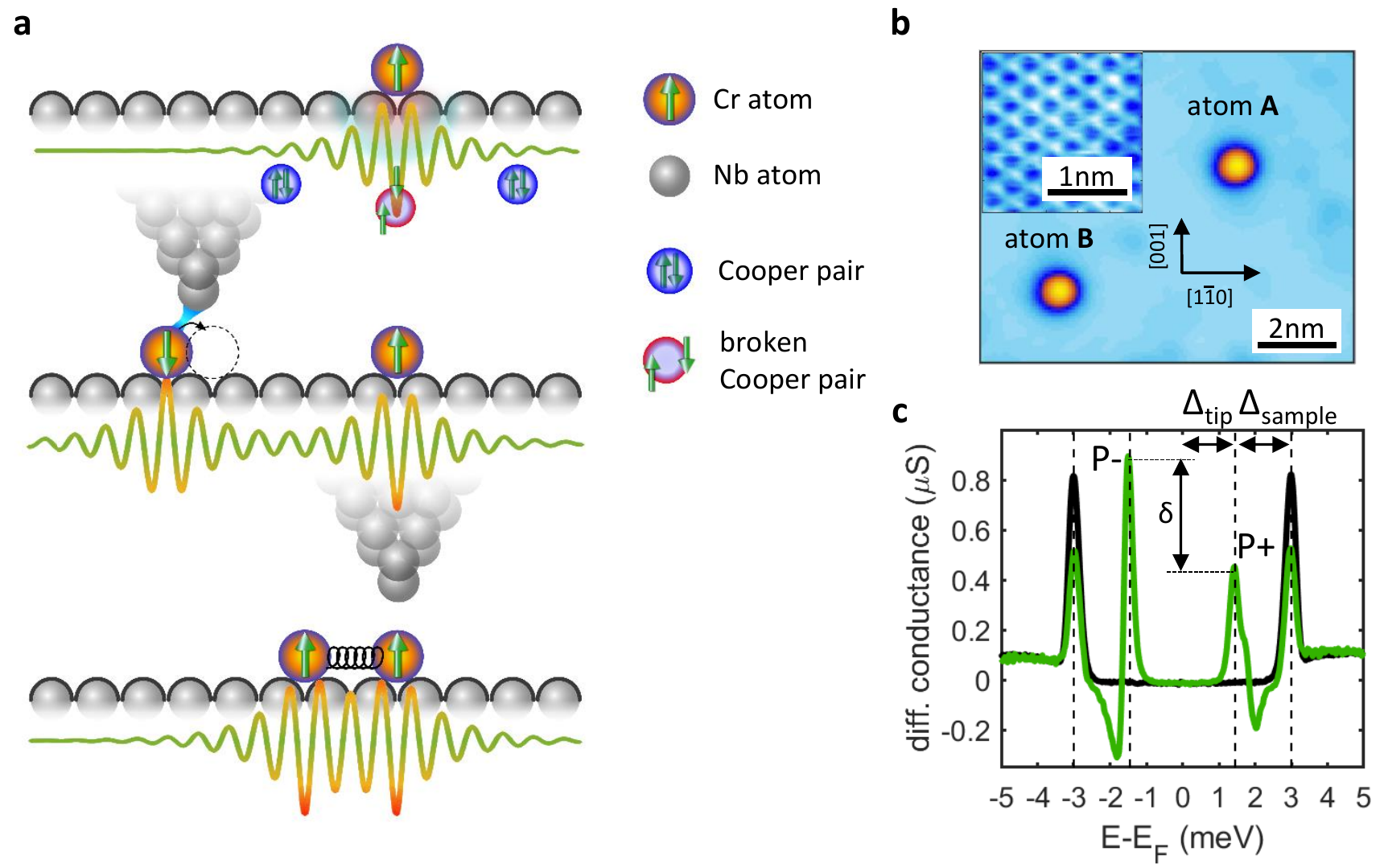}
	\caption{\textbf{Experimental lineup. a} Schematic illustration of coupling between local spins. Top: A single Cr magnetic moment interacts with the superconducting condensate and breaks locally Cooper pairs leading to the observed in-gap states. The YSR wave function spatially decays away from the impurity. Center: A second Cr atom is dragged closer by the STM tip. The YSR wave functions overlap leads to a detectable indirect coupling. Bottom: Two Cr atoms in next-nearest neighbor position. \textbf{b} Topographic image of the surface area used for the experiment showing Cr adatoms on the Nb(110) surface. Arrows indicate the crystallographic directions. Scan parameters: \SI{-5}{mV}; \SI{500}{pA}. Inset: Atomically resolved image of the clean Nb surface. Scan parameters: \SI{5}{mV}; \SI{5}{nA}.    	\textbf{c} Scanning tunneling spectra acquired with a Nb tip on clean Nb (black) and on top of an isolated Cr atom (green). Dashed lines indicate the superconducting gap size of the tip \protect$\Delta_\mathrm{tip}$ and sample \protect$\Delta_\mathrm{sample}$.} 
	\label{Figure1}
\end{figure*}

Figure \ref{Figure1}a schematically illustrates how  local magnetic moments can couple through a superconducting condensate. 
An isolated magnetic adatom (upper panel) induces the so-called Yu-Shiba-Rusinov (YSR) quasi particle state \cite{Yu.1965,Shiba.1968,Rusinov.1969,Yazdani.1997}  characterized by a spatially decaying wave function \cite{Menard.2015,Ruby.2018,PhysRevLett.120.167001,nanolett.7b05050,Kim.2020}. 
When a second magnetic atom is brought close enough (middle panel), the local spins start coupling to each other, with the interaction being mediated by the substrate. 
By progressively reducing their distance (bottom panel), the direct magnetic exchange becomes stronger and finally prevails over substrate mediated interactions.

Overall, this is expected to result in a rich phase diagram where both the sign and the strength of the interactions can be tuned on demand by controllably acting on distance between adatoms and crystallographic directions, as proposed in Ref. \onlinecite{PhysRevB.73.224511,PhysRevB.67.020502}.

As a prototypical system to scrutinize these aspects, we focus on Cr adatoms deposited on the (110) surface of Nb (see Methods for details on sample preparation). 
Niobium offers several advantages \cite{PhysRevB.102.174504,Schneidereabd7302,Kuester2021}. 
Its wide superconducting gap ($2 \Delta$ = \SI{3.05}{\milli\electronvolt} with $T_\mathrm{c}$ = \SI{9.2}{\kelvin}) allows to spectroscopically disentangle low-energy magnetic signatures from thermal broadening effects. 
Moreover, contrary to other superconductors such as Pb, its high cohesive energy makes it suitable for single-atom manipulation techniques \cite{Odobesko.2019,PhysRevB.102.174504,Schneidereabd7302}. 
Finally, its highly anisotropic Fermi surface \cite{PhysRevB.102.174502} (see also Fig.~\ref{Figure2}c) allows us to investigate the emergence of directional-dependent coupling between magnetic moments, which is expected to enhance the tunability. 

Figure \ref{Figure1}b shows a clean Nb area with the inset corresponding to the atomically resolved lattice. Only two Cr adatoms are present in the area, referred to as atom A and B. 
As described in the following, this allows to detect their interaction ruling out spurious effects such as impurities or the influence of other adatoms in the nearby.

Figure \ref{Figure1}c reports the STS spectrum acquired by positioning the tip on top of an isolated Cr atom (green line) and over the bare superconductor (black line). 
A Nb terminated tip has been used to enhance the energy resolution, resulting in the typical convoluted spectrum of tip and sample superconducting energy gap, $\Delta_\mathrm{tip}$ and $\Delta_\mathrm{sample}$ (see Supplementary Note 1). 
A rich spectroscopic scenario is visible for the Cr adatom, with several peaks emerging within the superconducting gap. 
These peaks are direct fingerprints of magnetic impurity-superconductor interactions, with magnetic moments locally breaking Cooper pairs\cite{Bardeen.1957} and inducing the YSR quasi particle resonances residing inside the superconducting energy gap~\cite{Yu.1965,Shiba.1968,Rusinov.1969,Yazdani.1997}. YSR states always appear in pairs that are energetically particle-hole symmetric with respect to the Fermi level, with the energy position $E_\textrm{YSR}$ being determined by the coupling strength between the magnetic impurity and the  superconducting condensate. Their orbital character can be visualized by spatially mapping their wave function distribution as shown in Supplementary Note 2  \cite{Ruby.2016,CRC2017,CRC2017}. 

The YSR pairs are generally characterized by a strong spectral weight asymmetry, the higher intensity in the occupied or unoccupied states being directly linked to a spin-screened and free-spin ground state, respectively. The crossing of YSR pairs through zero marks a first order quantum phase transition (QPT) between the two different regimes\cite{Franke.2011,Farinacci.2018,Huang.2020,Salkola1997}.
In the present case, the d$_{z^2}$-derived YSR state dominates the scene, showing a significant spectral weight asymmetry $\delta$ with higher intensity below the Fermi level (see Figure~\ref{Figure1}).
Note that the d$_{z^2}$-derived YSR pair is  very close to zero energy ($\pm\Delta_\mathrm{tip}$ in our case because of the use of superconducting tips). For strong interactions, occurring for adatoms placed at short distances, electron and hole components of the YSR pair can significantly shift becoming well-separated. The resulting energy shift $\Delta E_\textrm{YSR}$ can be directly evaluated based on the shift of their peak position with respect to the single adatom case. However, when the distance between the adatoms is increased, their interaction and consequently $\Delta E_\textrm{YSR}$ become progressively weaker. In such a regime, any small energy shift $\Delta E_\textrm{YSR}$  results in a substantial transfer of spectral weight between the positive ($E \simeq +\Delta_\mathrm{tip}$) and the negative ($E \simeq-\Delta_\mathrm{tip}$) peak. 
The change in peak height $\Delta\delta(\VEC{r})$ as function of the interatomic distance $r$ can thus be directly linked to the energy shift of the YSR pairs, i.e. $\Delta\delta(\VEC{r})\sim\Delta E_\textrm{YSR}$ (see Supplementary Note 4), offering a very sensitive measurement protocol to sense both the strength as well as the sign of the shift induced by the interaction between magnetic adatoms once YSR pairs are close to the QPT. These two different regimes are analyzed in Figure 2 and Figure 3, respectively.




\subsection{Directional dependence of short distant coupled spins.}

\begin{figure*}
        \renewcommand{\figurename}{Figure}
	\centering
	\includegraphics[width=.7\textwidth]{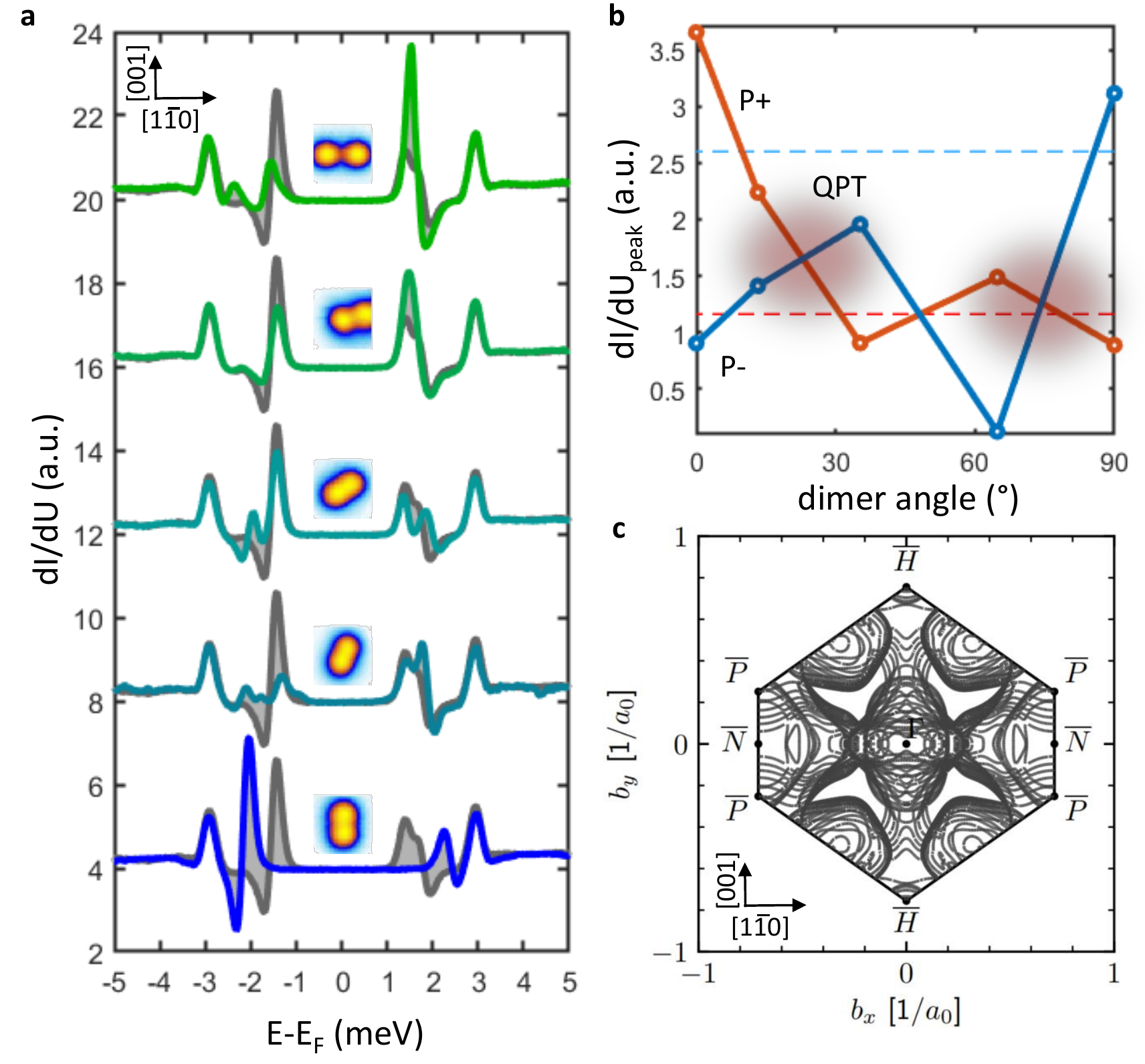}
	\caption{\textbf{Directional dependence. a} STS spectra taken on atom A for different positions of atom B. 
	STS on the isolated reference atom is shown in grey. 
	Curves are vertically shifted for clarity. 
	Insets: topography images $2\times2 \, \si{\nano\meter}$ of the corresponding Cr dimers. 
	\textbf{b} Maximum intensity of the in-gap peaks in the positive (red) and negative (blue) energies with respect to the Fermi level as a function of the dimer angle where \SI{0}{\degree} means parallel to the \protect$[1\overline{1}0]$ direction. 
	Transfer of spectral weight between positive and negative energies with highlighted quantum phase transition. 
	\textbf{c} Fermi surface of Nb(110) surface.}
	\label{Figure2}
\end{figure*}

Figure \ref{Figure2} compares Cr dimers aligned along 5 different directions while keeping them at a small distance, i.e.~below \SI{1}{\nano\meter}. 
Note that none of the cases corresponds to the first neighbor configuration. The corresponding $\mathrm{d}I/\mathrm{d}U$ signals are shown in Figure \ref{Figure2}a. The spectrum obtained on an isolated Cr adatom is provided as reference (gray line).
While the $[001]$ direction ( \SI{90}{\degree}) shows a negative energy shift compared to the single adatom case,  all other directions are characterized by a positive shift, in agreement with our theoretical calculations (see below). Coupled YSR states are also characterized by a distinct spatial distribution compared to the isolated adatom case, as illustrated in Supplementary Note 3.


Figure \ref{Figure2}b allows us to analyse these aspects more into details by showing the evolution of spectral intensity for the YSR pair dominating the scene within the superconducting gap (see Fig. \ref{Figure1}b). A significant transfer of spectral weight from hole- (red line) to electron-like (blue line) YSR states is visible by progressively increasing the angle between substrate and magnetic dimers. At specific angles, the lines are crossing each other, revealing the occurrence of a quantum phase transition from a free-spin (\SI{0}{\degree}, direction $[1\overline{1}0]$) into a screen-spin regime (\SI{90}{\degree}, direction $[001]$). These results  demonstrate the existence of highly anisotropic coupling between the magnetic impurities. The signature of the anisotropy of the substrate electronic structure can be grasped from the Fermi surface as calculated from ab-initio (see Methods) and illustrated in Figure \ref{Figure2}c. These effects are found to be remarkably strong: they allow to tune not only the strength but also the sign of energy shift induced by the adatom-adatom interaction by properly acting on the crystallographic direction. 


\subsection{Distance and directional dependence of indirectly coupled spins.}
\begin{figure*}
        \renewcommand{\figurename}{Figure}
	\centering
	\includegraphics[width=.5\textwidth]{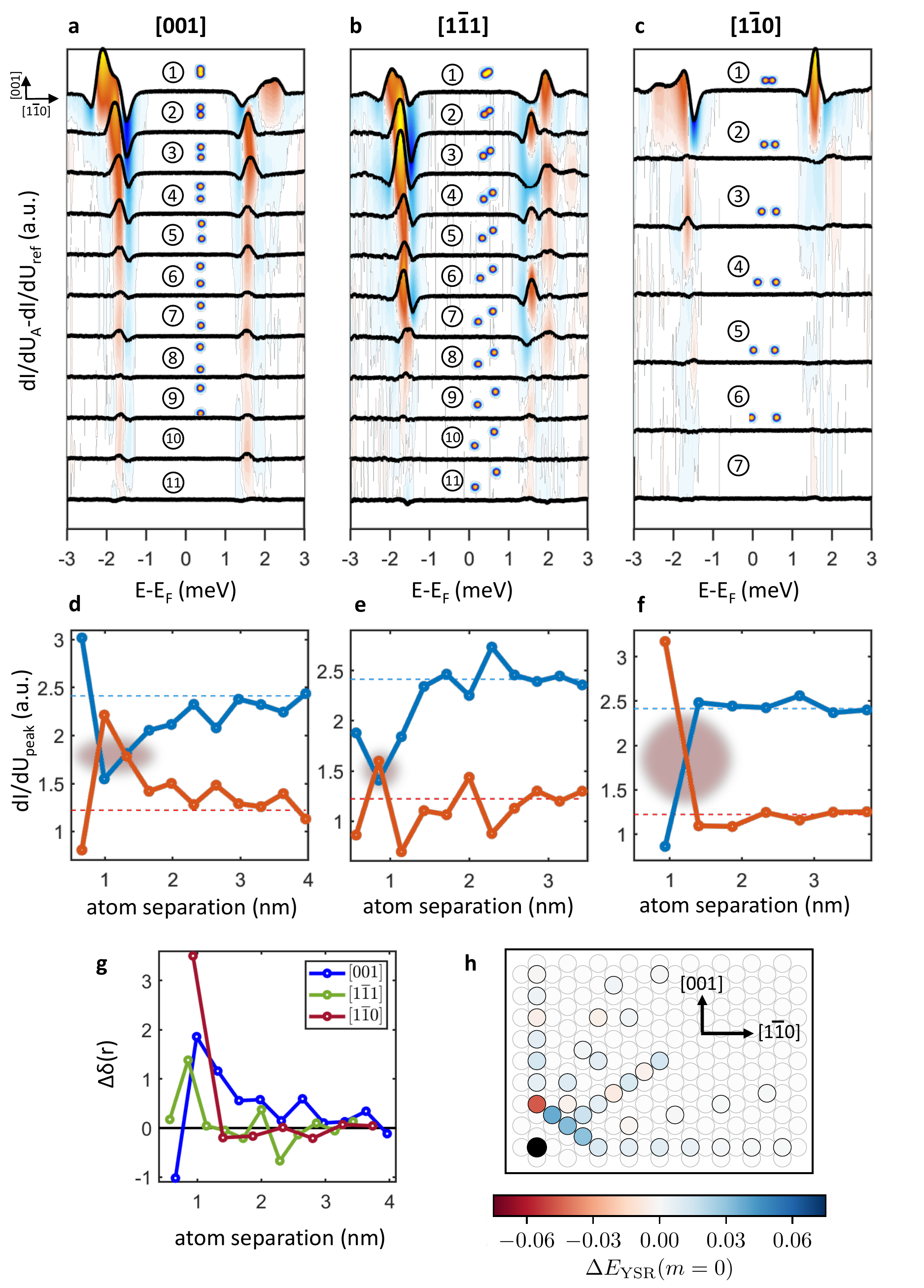}
	\caption{\textbf{Distance dependence of the YSR energy spectrum. a-c} STS taken on atom A subtracted by the signal on the isolated Cr atom. Spectra are shown for three different directions and several distances, the number in each panel corresponding to the number of discrete steps.   
	\textbf{d-f} In-gap peak maximum for positive (red) and negative (blue) energies as a function of the distance between the adatoms. The dashed lines show the isolated atom intensities. Red shaded areas highlight the occurrence of a quantum phase transition from a free-spin to a screened-spin regime. 
	\textbf{g} Comparison of spectral weight redistribution along distinct directions and at different distances. Zero corresponds to the isolated adatom case. 
	\textbf{h} First-principles calculations of the shift of the YSR energies of the $z^2$-derived orbital with respect to the single adatom YSR energies. The black dot illustrates the fixed atom A, while the colored positions represent atom B and therefore the formed dimer configuration.
	Adsorption site positions on the Nb lattice corresponding to the cases measured experimentally. The color code indicates the calculated shift in YSR energy for the respective Cr dimers.
	}
	\label{Figure3}
\end{figure*}

To investigate the evolution of the interaction over larger distances, we performed a distance dependence study of the coupling between local spins along three distinct crystallographic directions, i.e. $[001]$, $[1\overline{1}1]$ and $[1\overline{1}0]$. The distance between two atoms has been progressively increased step-by-step, with each step being defined by the smallest possible vector compatible with the underlying lattice periodicity, i.e. the vector connecting two subsequent adorption sites. 

Figure \ref{Figure3}a-c reports the spectra obtained for each configuration of the adatoms subtracted by the spectrum obtained on the isolated atom (reference spectra are reported in Supplementary Note 5). This procedure allows to unambiguously detect and quantitatively analyse the effects induced by the coupling (see Supplementary Note 4).  In line with expectations, our data reveal that the coupling becomes progressively stronger by reducing the distance between the adatoms.  However, as described in the following, the coupling is found to persist at distances significantly larger than those observed in metallic systems \cite{Zhou.2010,Khajetoorians.2012,Khajetoorians2016}. Interestingly, the data obtained at larger distances reveal also that the  decay length of the interaction is highly anisotropic. 
While along the $[001]$ and $[1\overline{1}1]$ direction a clear coupling between the magnetic adatoms can be detected up to distances of approximately \SI{4}{\nano\meter}, the interaction decays much faster along the $[1\overline{1}0]$ direction, where the interaction is strongly suppressed already at approximately \SI{1.4}{\nano\meter}. 
The different crystallographic direction behaviour can be traced back to the anisotropy of the Fermi surface. Good nesting vectors between flat parts of the Fermi surface can have a focusing effect on the wave function of quasi particles, significantly enhancing the coupling strength over larger distances \cite{Weismann.2009,Lounis2011,Bouhassoune2014}. Although the Fermi surface is rather complex in the present case, good nesting conditions are expected to manifest along the $[001]$ direction, in line with our experimental observations (see Fig.~\ref{Figure2}c).


A careful inspection of Figure \ref{Figure3} additionally reveals the existence of an oscillatory behavior of the coupling strength which is superimposed to the decay. This is clearly visible along the $[001]$ and $[1\overline{1}1]$ directions, while it becomes more subtle along the $[1\overline{1}0]$ direction.

These oscillations are further highlighted in Figure \ref{Figure3}d-e, where the in-gap peak intensities are plotted separately for negative and positive energies. The comparison between electron- and hole-like states clearly reveal that their distance-dependence trends are anticorrelated. This observation confirms the transfer of spectral weight within the YSR pair. 
Note that a distance-dependent quantum phase transition from a free-spin to a screened-spin regime takes place along all directions, highlighted by a red cloud in Fig.\ref{Figure3}d-e.

Figure \ref{Figure3}g directly compares the change of  the peaks asymmetry with respect to the isolated adatom $\Delta\delta$ as a function of the atom separation $r$. Positive and negative values correspond to a shift $\Delta E_\mathrm{YSR}$ to positive and negative energies, respectively. Although the spatial resolution imposed by the lattice impedes a high sampling rate, the curves suggest a similar oscillation period for the different directions (see Supplementary Notes 6 and 7). 

To explore the origin of the observed behavior, we calculated the change of the YSR energy of the $z^2$-derived orbital of the dimer atoms with respect to the YSR energy of the isolated adatom, $\Delta E_\mathrm{YSR}$, using first principles and an effective Hamiltonian construction (see Methods and Supplementary Note 8 and 9), which is shown in Figure \ref{Figure3}h using a spatial map for all discussed dimer configurations.
In good agreement with the experimental observations, the energy shift is strongest for the next-nearest neighbor dimers along the two high-symmetry directions $[001]$ and $[1\overline{1}0]$ and significantly weaker for the directions in between.
The simulations also shed light on the fundamental physical origins determining the strength of the YSR energy shifts: (i) the nature of the magnetic coupling begin ferromagnetic or anti-ferromagnetic, (ii) the strength of the direct electronic coupling between the dimer atoms, and (iii) the strength of the indirect superconductor-mediated electronic and magnetic coupling between the dimer atoms. 
Analytic model calculations based on a simplified Alexander-Anderson model~\cite{Oswald1985} (see Supplementary Note 9) predict distinct shifts of the YSR energies depending on ferromagnetic or antiferromagnetic interactions. These results from the influence of the magnetic coupling on the formation of the electronic bonding and anti-bonding states~\cite{PhysRevB.61.14810,Yao2014}. 
However, spin-orbit coupling can lift this degeneracy resulting in split YSR states also for AFM coupled spins, as shown in Ref. \onlinecite{BSR2021} and Supplementary Note 3. 
The experimentally unveiled magnetic structures reported in Supplementary Note 10 and the more advanced theoretical model based on ab-initio parameters (leading to Fig. 3h) demonstrate that the sign of the energy shift and the occurrence of a QPT can not be trivially linked to the sign of the  magnetic exchange interaction. Indeed, in agreement with theoretical calculations, ferromagnetic order has been found for adatoms coupled along the $[111]$ direction at the next nearest-neighbor distance, while those oriented along $[001]$ and $[1\overline{1}0]$ both show antiferromagnetically coupled adatoms, despite being characterized by opposite energy shift of the YSR pairs, as correctly captured in our calculations reported in Fig. 3h.



\subsection{Indirect coupling between spins in the metallic regime.}
\begin{figure*}
        \renewcommand{\figurename}{Figure}
	\centering
	\includegraphics[width=.9\textwidth]{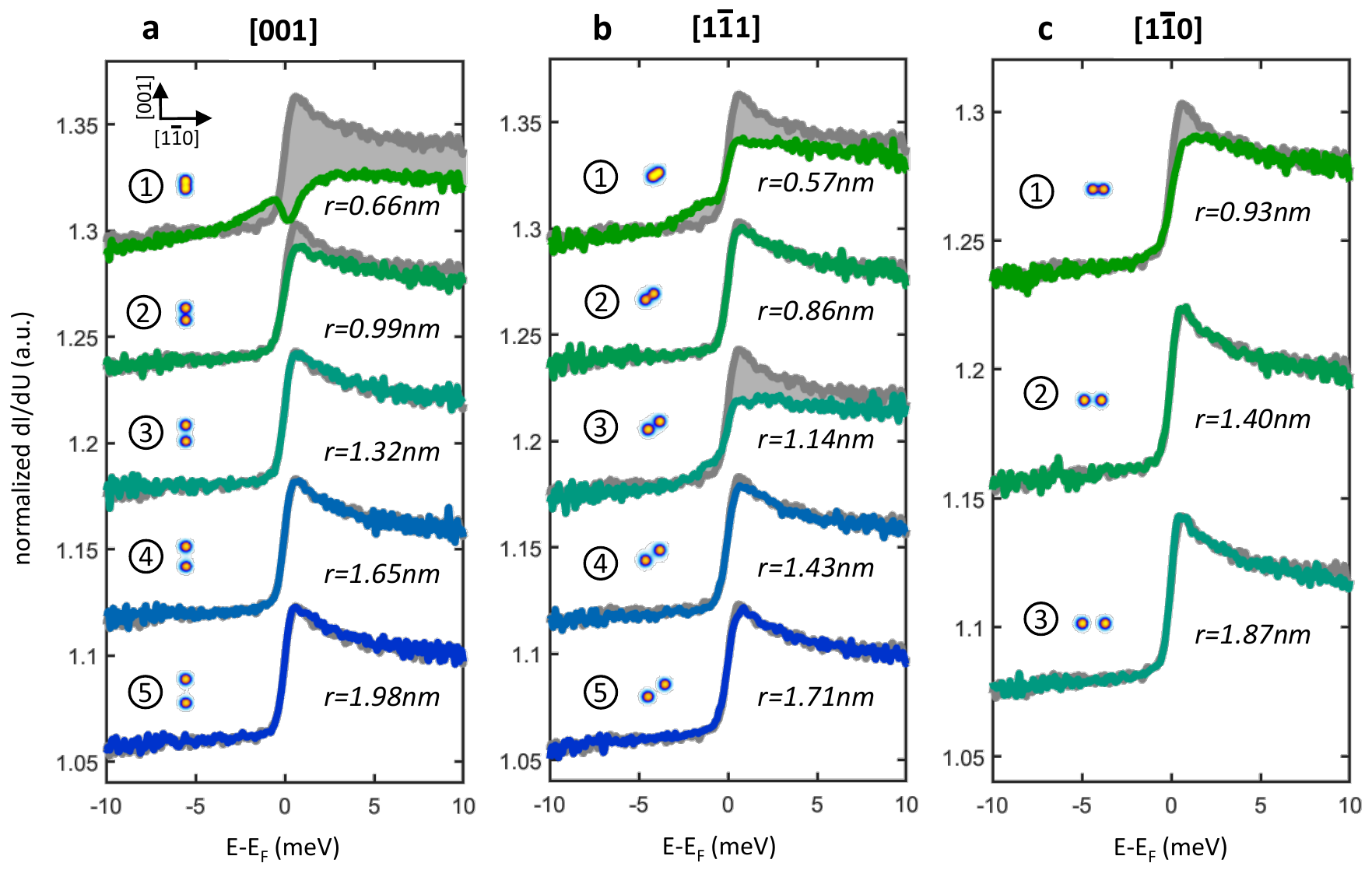}
	\caption{\textbf{Spectroscopy in the metallic regime. a-c}. \protect$\mathrm{d}I/\mathrm{d}U$ signals over atom A with increasing distance of atom B starting with step 1 at next-nearest neighbor positions for three different crystallographic directions. The signal of the isolated Cr adatom is shown in grey for each spectrum as a reference. An external out of plane magnetic field of \SI{1}{\tesla} was applied to fully suppress superconductivity. Stabilization parameters: \SI{-10}{\milli\volt}; \SI{3}{\nano\ampere}. Insets: topographic images of the corresponding dimer shown in equal scale.}
	\label{Figure4}
\end{figure*}

To corroborate our findings, we compare the coupling between adatoms in the superconducting regime with the one observed when the interaction is mediated by conventional conduction electrons. To this end, the substrate has been driven into the metallic regime by applying a magnetic field of \SI{1}{\tesla} perpendicular to the sample surface.  As described in the following, this allows to confirm the existence of highly anisotropic indirect interactions as well as to highlight to the existence of a distinct contribution to the magnetic coupling which originates from the superconducting condensate.

Figure \ref{Figure4}a-c reports the spectra acquired over the Cr adatoms by progressively changing their relative distance along each one of the three distinct directions scrutinized in the superconducting regime, i.e. $[001]$, $[1\overline{1}1]$, and $[1\overline{1}0]$. All spectra are overlapped to a reference spectrum obtained for an isolated Cr atom (grey line). The spectrum of the Cr single adatom is characterized by a sharp step-like signature at the Fermi level which has recently been demonstrated being directly linked to its magnetic ground state \cite{Kuester2021}. Changes in this spectral shape can thus be used to reveal the existence of magnetic interactions between the adatoms.

The direct comparison of the spectra acquired along the three distinct directions when adatoms are at the closest distance (step 1 in Figure \ref{Figure4}a-c) confirms the existence of a highly anisotropic indirect interaction. This effect can be effectively visualized by comparing the gray areas in panels (a-c), which highlight the crystallographic direction-dependent changes with respect to the single adatom case. The largest spectral change is observed along the $[001]$ direction. Along this direction, the step-like feature characteristic of the single adatom is significantly suppressed and broadened, signaling a strong magnetic interaction between the adatoms. The spectral change is weaker along the $[1\overline{1}1]$ direction and it becomes almost negligible along the $[1\overline{1}0]$ direction. Note that the distance between the adatoms is smaller along the $[1\overline{1}1]$ direction ($r = \SI{0.57}{\nano\meter}$) compared to the $[001]$ direction ($r = \SI{0.66}{\nano\meter}$). If the effect would be distance-dependent,  the spectral change along the $[1\overline{1}1]$ direction should be stronger than the one along the $[001]$ direction. Our data clearly reveal the opposite. This allows to identify the highly anisotropic Fermi surface as the origin of the anisotropic coupling, supporting the conclusions drawn from the analysis of the superconducting regime.

Finally, we investigate the distance-dependence behavior of conduction electrons-mediated interactions. Figure \ref{Figure4} (a-c) evidence that along the $[001]$ direction, where the interaction at short distances is the strongest, the spectrum is perfectly overlapping the isolated adatom case already at $r = \SI{1.32}{\nano\meter}$. A slower decay is observed along the $[1\overline{1}1]$ direction, where the  difference with respect to the isolated adatom case shows an oscillatory on-off behavior depending on the odd or even number of steps between the adatoms, respectively. An oscillatory behavior along this crystallographic direction was also revealed in the superconducting regime (see Figure \ref{Figure3} e and g). 
Finally, along the $[1\overline{1}0]$ direction, the difference with respect to the single adatom vanishes already at step 2, an observation in line with the very rapid spatial decay of the interaction between magnetic moments observed also in the superconducting regime.

The decay length in the metallic regime is in line with earlier STM studies on other metallic systems \cite{Zhou.2010,Khajetoorians.2012,Khajetoorians2016} and significantly faster than the one observed in the superconducting regime, where interactions between adatoms could be detected up to \SI{4}{\nano\meter} (see Supplementary Note 6 and 7). All together, these observations could be originating from the decisive role of the superconducting condensate in mediating interactions over long distances, an effect predicted decades ago but which always escaped experimental verification. 


\section*{Discussion}

In conclusion, we demonstrated the emergence of long range, crystallographic direction dependent, and highly tunable superconducting condensate-mediated coupling between local spins. 
We could directly link the strong crystallographic direction dependence of the coupling to the highly anisotropic Fermi surface, which can be used to controllably cross the QPT between screen-spin and free-spin regime by changing the alignment of the atom pair with respect to the underlying lattice with atomic precision. 
Longer range coupling effects between magnetic impurities can be observed along the $[001]$ and $[1\overline{1}1]$ directions compared to $[1\overline{1}0]$, an enhancement which is potentially enabled by the optimal Fermi surface nesting conditions significantly strengthening substrate mediated magnetic interactions. 
Our results unveil, in an extremely simple material platform,  the existence of a very rich phase diagram which can be used for tuning magnetic interactions mediated by superconducting condensates on demand. The simultaneous fulfilling of the following requirements: (i) magnetic impurities hosting deep YSR states, (ii) indirect coupling between adatoms taking place at large distances and (iii) the possibility to cross the QPT upon hybridization \cite{PhysRevB.92.125422}, create ideal conditions for experimentally scrutinizing the emergence of topologically superconductivity in dilute spin systems \cite{PhysRevB.88.155420} going beyond the nearest neighbors and ferromagnetically coupled spin chains explored so far \cite{Nadj-Perge602,PhysRevLett.115.197204,SBP2021}. We envision our results to be applicable to more complex systems and phenomena, creating precise guidelines for the rational design of exotic phenomena emerging at magnetic-superconducting interfaces \cite{steiner2021quantum,mishra2021yushibarusinov}.

\begin{methods}

\subsection{Sample and tip preparation}
The bulk single crystal Nb(110) substrate was cleaned in UHV by repeatedly heating the surface to \SI{2300}{\kelvin} in \SI{12}{\second}~\cite{Odobesko.2019}. Cr adatoms were deposited in-situ and with the substrate below a temperature of \SI{15}{\kelvin}. All atoms were found to be adsorbed in the hollow site of the Nb(110) surface~\cite{Kuester2021}. Measurements were taken in UHV at \SI{600}{\milli\kelvin} using the Tribus STM head (Scienta Omicron). Cr adatoms could be moved with atomic precision by approaching them with the STM tip in constant current mode and a setpoint of \SI{-5}{\milli\volt}; \SI{70}{\nano\ampere}. $\mathrm{d}I/\mathrm{d}U$ spectra were taken with the standard lock-in technique. For measurements in the normal metallic regime we used a tungsten tip characterized on Ag(111) and applied a magnetic field of $>$\SI{1}{\tesla}. For measurements in the superconducting regime, we used a superconducting Nb microtip obtained by deep indentations into the Nb single crystal. As described in Supplementary Note 1, this allows to improve the energy resolution.

\subsection{Measurement Protocol}

Throughout the distance dependence experiments, atom A was not moved from its initial position while atom B was subsequently positioned in a controlled way to specific hollow site positions, so that all changes seen on atom A are exclusively triggered by movements of atom B, therefore all spectra shown were taken on atom A.
Both at the start and the end of the experiment, the very same configuration shown in Figure \ref{Figure1}a was arranged (atom distance $|\VEC{r}_\textrm{ref}|=\SI{5.6}{\nano\meter}$) and measured to rule out any tip changes during the multiple atom manipulations required. 
An exactly identical $\mathrm{d}I/\mathrm{d}U$ signal could be acquired before and after and was then used as the reference for all other spectra.

\subsection{First-principles calculations}
Our first-principles approach is implemented in the framework of the scalar-relativistic full-electron Korringa-Kohn-Rostoker (KKR) Green function augmented self-consistently with spin-orbit interaction~\cite{Papanikolaou:2002,Bauer:2014}.
The method is based on multiple-scattering theory allowing an embedding scheme, which is versatile for the treatment of nanostructures in real 
space. 
The full charge density is computed within the atomic-sphere approximation (ASA) and local spin density approximation (LSDA) is employed for the evaluation of the exchange-correlation potential~\cite{Vosko:1980}. 
We assume an angular momentum cutoff at $\ell_{\text{max}} = 3$ for the orbital expansion of the Green function and when extracting the local density of states a k-mesh of $150 \times 150$ is considered.   
The Nb(110) surface is modelled by slab containing 22 layers enclosed by two vacuum regions with a thickness of \SI{9.33}{\angstrom} each.
The Cr adatoms are placed on the hollow stacking site relaxed towards the surface by \SI{20}{\percent} of the inter-layer distance of the underlying Nb(110) surface, which was shown to be the energetically favoured stacking in Ref.~\cite{Kuester2021}.
\\
The magnetic exchange interactions in the dimers were obtained using the magnetic force theorem in the frozen-potential approximation and the infinitesimal rotation method \cite{Liechtenstein1987,Ebert2009}.
The on-site magnetic anisotropy of the Cr atoms in all dimers is estimated by the one of the isolated Cr adatom, which is obtained from the method of constraining fields \cite{Brinker2019}.
Furhter information can be found in Supplementary Note 7.
\\
The change of the YSR energies ($\Delta E_\mathrm{YSR}$) is estimated using a tight-binding model with parameters obtained from density functional theory and an effective Hamiltonian construction (see Ref. \cite{Schneider2020} and Supplementary Note 8).

\end{methods}

\begin{addendum}


\item[Competing Interests] The authors declare that they have no competing interests. 

\item[Data and materials  availability] All data needed to evaluate the conclusions in the paper are present in the paper and/or the supplementary materials. Additional data related to this paper may be requested from the authors.  The KKR Green function code that supports the findings of this study is available from the corresponding author on reasonable request.
\end{addendum}

\newpage

\section*{References}
\bibliographystyle{naturemag}
\bibliography{bibliography}

\end{document}